\documentclass[twocolumn,showpacs,preprintnumbers]{revtex4}%
\usepackage{amssymb}
\usepackage{amsmath}
\usepackage{graphicx}
\usepackage{dcolumn}
\usepackage{bm}
\usepackage{amsfonts}%
\begin{document}
\title{Time Resolved Phase Space Tomography of an Optomechanical Cavity}
\author{Oren Suchoi}
\author{Keren Shlomi}
\author{Lior Ella}
\author{Eyal Buks}
\affiliation{Department of Electrical Engineering, Technion, Haifa 32000 Israel}
\date{\today }

\begin{abstract}
We experimentally study the phase space distribution (PSD) of a mechanical
resonator that is simultaneously coupled to two electromagnetic cavities. The
first one, operating in the microwave band, is employed for inducing either
cooling or self-excited oscillation, whereas the second one, operating in the
optical band, is used for displacement detection. A tomography technique is
employed for extracting the PSD from the signal reflected by the optical
cavity. Measurements of PSD are performed in steady state near the threshold
of self-excited oscillation while sweeping the microwave cavity detuning. In
addition, we monitor the time evolution of the transitions from an
optomechanically cooled state to a state of self excited oscillation. This
transition is induced by abruptly switching the microwave driving frequency
from the red-detuned region to the blue-detuned one. The experimental results
are compared with theoretical predictions that are obtained by solving the
Fokker-Planck equation. The feasibility of generating quantum superposition
states in the system under study is briefly discussed.

\end{abstract}
\maketitle





Optomechanical cavities are currently a subject of intense basic and applied
study \cite{Braginsky_653,Braginsky&Manukin_67,
Hane_179,Gigan_67,Metzger_1002,Kippenberg_1172,Favero_104101,Marquardt2009,Marquardt_0905_0566,girvin_Trend}%
. These devices can be employed in various sensing \cite{Rugar1989,
Arcizet2006a, Forstner2012,Weig2013,Braginsky_QM,Clerk_1155} and photonics
applications \cite{Lyshevski&Lyshevski_03,Stokes_et_al_90,
Hossein_Zadeh_276,Wu_et_al_06,
MattEichenfield2007,Bahl2011,Flowers-Jacobs2012,Bagheri_726,Zhou_179}.
Moreover, optomechanical cavities may allow experimental study of the
crossover from classical to quantum mechanics \cite{Thompson_72,
Meystre2013,Kimble_et_al_01, Carmon_et_al_05, Arcizet_et_al_06, Gigan_67,
Jayich_et_al_08, Schliesser_et_al_08, Genes_et_al_08,
Teufel_et_al_10,Poot_273,Tang_1404_5574} and the observation of macroscopic
quantum behavior in mechanical systems
\cite{Teufel_204,OConnell_697,Kimble_et_al_01,Genes_et_al_08,Teufel_et_al_10,Rodrigues_053601,Qian_1112_6200,Palomaki_710,He_052121,Pikovski_393,Poot_273,Meaney_056202,Walter_1307_7044,Lorch_011015,Galland_1312_4303,Kiesewetter_1312_6474,Bahrami_1402_5421,Farace_1306_1142,Xu_063819,Weinstein_1404_3242,Xu_1404_3726}%
. When the finesse of the optical cavity that is employed for constructing the
optomechanical cavity is sufficiently high, the coupling to the mechanical
resonator that serves as a vibrating mirror is typically dominated by the
effect of radiation pressure \cite{Kippenberg_et_al_05,Rokhsari2005,
Arcizet2006,Gigan_et_al_06,Cooling_Kleckner06,
Kippenberg_1172,Corbitt_150802,Schliesser_243905,Nichols_307}. On the other
hand, bolometric effects can contribute to the optomechanical coupling when
optical absorption by the vibrating mirror \cite{Aubin_et_al_04,
Marquardt_103901, Paternostro_et_al_06, Liberato_et_al_10_PRA} is significant
\cite{Metzger_1002, Jourdan_et_al_08,Marino&Marin_10,Marino&Marin2011PRE,
Metzger_133903, Restrepo_860, Liberato_et_al_10,Marquardt_103901,
Paternostro_et_al_06,Yuvaraj_430,Zaitsev_046605,Zaitsev_1589}. In recent years
a variety of cavity optomechanical systems have been constructed and studied
\cite{Hane_179,Kippenberg_1172,Corbitt_S675,Corbitt_150802,Metzger_1002,Gigan_67,Arcizet_et_al_06,Kleckner_75,Favero_104101,Regal_555,Carmon_223902,Schliesser_243905,Thompson_72,Chan_89,Teufel_359,Teufel_204,OConnell_697,Groeblacher_181104}%
, and phenomena such as mode cooling
\cite{Teufel_359,Teufel_204,Groblacher_485,Schliesser_509,Chan_89},
self-excited oscillation
\cite{Hane_179,Kim_1454225,Aubin_1018,Carmon_223902,Marquardt_103901,Corbitt_021802,Carmon_123901,Metzger_133903,Regal_555}
and optically induced transparency
\cite{Weis_1520,Karuza_013804,Safavi-Naeini_69,Ojanen_1402_6929} have been investigated.%

\begin{figure}
[ptb]
\begin{center}
\includegraphics[
height=4.3941in,
width=3.2396in
]%
{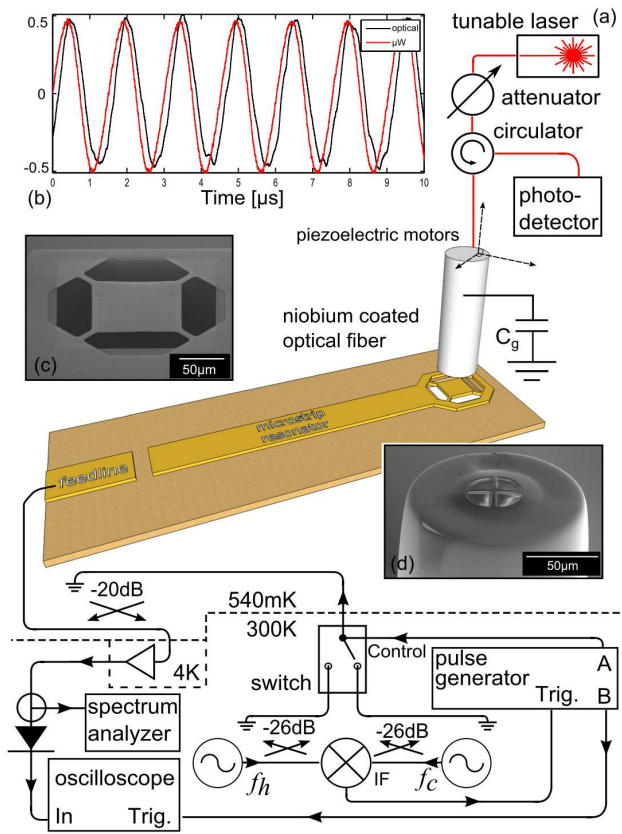}%
\caption{(color online) Experimental Setup. The dual optomechanical cavity is
seen in panel (a). The Microwave cavity is a superconducting microstrip made
of aluminum over a high resistivity silicon wafer coated with a
$100\operatorname{nm}$ thick SiN layer. The optomechanical coupling is
generated using a Nb coated optical fiber that is positioned at sub-micron
distance from the trampoline. The optical setup (seen above the sample) allows
using the optical cavity for displacement detection, whereas the microwave
setup (seen below the sample) is employed for exciting the microwave cavity
and measuring its response. For the measurements of PSD in steady state, a
single microwave synthesizer is employed, whereas for the time-resolved
measurements, two synthesizers (tuned to frequencies $f_{\mathrm{c}}$, and
$f_{\mathrm{h}}$, respectively) are used together with a pulse generator in
order to ensure a smooth switching from cooling to heating. Panel (b) shows
the off reflected signals of both the optical and the microwave cavities, in
the region where the system exhibits self-excited oscillation in steady state.
The mechanical resonator at the end of the microstrip is seen in the electron
micrograph in panel (c). Several windows are opened in the Nb layer on the
fiber tip using FIB, as can be seen in panel (d).}%
\label{Fig setup and SEO}%
\end{center}
\end{figure}

In this work we experimentally study a hybrid system made of a single
mechanical resonator and two cavities, one operating in the microwave band and
the other in the optical band. We study self-excited oscillation induced by
driving the microwave cavity in the blue-detuned region. The technique of
state tomography is employed in order to construct the phase space
distribution (PSD) of the mechanical resonator \cite{Vanner_1406_1013}, whose
displacement is detected using the optical cavity. We begin by mapping out the
PSD in steady state near the threshold of self-excited oscillation. We then
employ time resolved measurements of PSD for both, studying the drifting in
time of the phase of self-excited oscillation, and for monitoring the
transition from cooling to self-excited oscillation. In the latter case, the
frequency of the driving signal that is injected into the microwave cavity is
abruptly switched from the red-detuned region to the blue-detuned one,
allowing thus the recording of the time evolution from an initial state, in
which the mechanical resonator is cooled down, to a final state, in which the
system undergoes self-excited oscillation. The possibility of employing such
abrupt switching for the creation of quantum superposition states is briefly discussed.

The experimental setup is schematically depicted in Fig.
\ref{Fig setup and SEO}. A photo-lithography process is used to pattern a
microwave microstrip cavity made of $200%
\operatorname{nm}%
$ thick aluminum on a high resistivity silicon wafer. At the open end of the
microstrip, a mechanical resonator in the shape of a $100\times100%
\operatorname{\mu m}%
^{2}$ trampoline supported by four beams is fabricated \cite{Zaitsev_046605}.
At the other end, the microwave cavity is weakly coupled to a feedline, which
guides both the injected and reflected microwave signals. Details of the
fabrication process can be found elsewhere \cite{Suchoi_1405_3467}.

The sample is mounted inside a closed Cu package, which is internally coated
with Nb. Measurements are performed in a dilution refrigerator operating at a
temperature of $0.54%
\operatorname{K}%
$ and in vacuum. The injected microwave signal is first attenuated by a $20$
dB directional coupler at the same temperature before entering the feedline.
The reflected microwave signal is amplified using a cryogenic amplifier at $4%
\operatorname{K}%
$, and measured in both the frequency domain (using a network analyzer and a
spectrum analyzer) and in the time domain (using a power diode connected to an
oscilloscope). The fundamental microwave cavity resonance frequency is
$f_{\mathrm{c}}=\omega_{\mathrm{c}}/2\pi=2.783%
\operatorname{GHz}%
$, the corresponding linear damping rate is $\gamma_{\mathrm{c}}=4.2%
\operatorname{MHz}%
$, the fundamental mechanical resonance frequency is $f_{\mathrm{m}}%
=\omega_{\mathrm{m}}/2\pi=662.7%
\operatorname{kHz}%
$ and the corresponding mechanical linear damping rate is $\gamma_{\mathrm{m}%
}=2.5%
\operatorname{Hz}%
$.

A single mode optical fiber coated with Nb is placed above the suspended
trampoline (see Fig. \ref{Fig setup and SEO}). In the presence of the coated
fiber, two optomechanical cavities are formed, one is a superconducting cavity
operating in the microwave band and the other is a fiber-based cavity
operating in the optical band. The coupling between the mechanical resonator
and the microwave cavity, originating by the capacitance between the coated
fiber and the suspended trampoline \cite{Suchoi_1405_3467}, is dominated by
the effect of radiation pressure, whereas bolometric effects are responsible
for the coupling between the mechanical resonator and the optical cavity. The
fact that both optomechanical cavities share the same mechanical resonator can
be exploited for inducing mechanically mediated coupling between microwave and
optical photons
\cite{Andrews_1310_5276,Jiang_1404_3928,Fong_1404_3427,Bochmann_122602,Yan_1405_6506,Wang_1406_7829,Tian_1407_3035,Clader_012324,Yin_1407_4938}%
.

A fiber Bragg grating (FBG) mirror and a focusing lens are made near the fiber
tip. To allow optical transmission, the core of the fiber at the tip is
exposed by etching the Nb coating using focused ion beam (FIB)
\cite{Suchoi_1405_3467}. A cryogenic piezoelectric 3-axis positioning system
having sub-nanometer resolution is employed for manipulating the position of
the optical fiber. The reflected signal off the optical cavity, which is
measured both by a spectrum analyzer and by an oscilloscope, is employed for
displacement detection. A tunable laser operating near the Bragg wavelength
$\lambda_{\mathrm{B}}=1545.8%
\operatorname{nm}%
$ of the FBG together with an external attenuator are employed to excite the
optical cavity. In order to avoid back-reaction effects originating by
bolometric effects \cite{Zaitsev_46605}, the input laser power feeding the
optical cavity is kept at sufficiently low level (below $60%
\operatorname{\mu W}%
$), allowing the utilization of the optical cavity for displacement detection
without any significant effect on the dynamics of other parts of the system.

Self-excited oscillation can be induced by injecting a monochromatic pump
signal into the feedline of the microwave cavity provided that the angular
frequency of the pump signal $\omega_{\mathrm{p}}$ is blue-detuned with
respect to the angular cavity resonance frequency $\omega_{\mathrm{c}}$, i.e.
provided that $\omega_{\mathrm{p}}>\omega_{\mathrm{c}}$, and provided that the
injected power $P_{\mathrm{p}}$ exceeds a threshold value. Panel (b) of Fig.
\ref{Fig setup and SEO} shows time traces of the off reflected signals of
both, the optical and the microwave cavities, in the region where the system
exhibits self-excited oscillation induced by a microwave monochromatic pump
tone having normalized detuning $d=\left(  \omega_{\mathrm{p}}-\omega
_{\mathrm{c}}\right)  /\omega_{\mathrm{m}}=0.7$ and power $P_{\mathrm{p}}=0.63%
\operatorname{\mu W}%
$. The off reflected optical signal is measured using a photo-detector, and
the off reflected microwave signal is measured using a power diode. The
relative phase between the two oscillating signals allows the direct
measurement of the retardation \cite{Braginsky_9906108} in the response of the
microwave cavity to mechanical oscillation (note that the retardation in the
response of the optical cavity is negligibly small).

The PSD provides an important insight on the effect of noise on the dynamics
of the system under study in both the classical and quantum regimes. Armour
and Rodrigues \cite{Armour_440} have calculated the Wigner quasi-probability
PSD for an optomechanical cavity for the case where the coupling between the
mechanical resonator and the cavity is dominated by the effect of radiation
pressure. They found that in the semiclassical approximation (in which
third-order derivative terms in the equation of motion for the Wigner function
are neglected) the master equation leads to Langevin equations corresponding
to the classical dynamics of the system. Furthermore, the assumption that the
dynamics of the mechanical resonator is slow on the time-scale of the cavity
dynamics allows employing the technique of adiabatic elimination in order to
derive an evolution equation for the complex amplitude $A$ of the mechanical
resonator, which is found to be given by%
\begin{equation}
\dot{A}+\left(  i\Omega_{\mathrm{eff}}+\Gamma_{\mathrm{eff}}\right)
A=\xi\left(  t\right)  \;, \label{A dot}%
\end{equation}
where both the effective angular resonance frequency detuning $\Omega
_{\mathrm{eff}}$ and the effective damping rate $\Gamma_{\mathrm{eff}}$ are
real even functions of $A_{r}\equiv\left\vert A\right\vert $. To second order
in $A_{r}$ they are expressed as $\Omega_{\mathrm{eff}}=\omega_{0}+\omega
_{2}A_{r}^{2}$ and $\Gamma_{\mathrm{eff}}=\gamma_{0}+\gamma_{2}A_{r}^{2}$. The
fluctuating term \cite{Risken_Fokker-Planck} $\xi\left(  t\right)  =\xi
_{x}\left(  t\right)  +i\xi_{y}\left(  t\right)  $, where both $\xi_{x}$ and
$\xi_{y}$ are real, represents white noise and the following is assumed to
hold: $\left\langle \xi_{x}\left(  t\right)  \xi_{x}\left(  t^{\prime}\right)
\right\rangle =\left\langle \xi_{y}\left(  t\right)  \xi_{y}\left(  t^{\prime
}\right)  \right\rangle =2\Theta\delta\left(  t-t^{\prime}\right)  $ and
$\left\langle \xi_{x}\left(  t\right)  \xi_{y}\left(  t^{\prime}\right)
\right\rangle =0$, where $\Theta=\gamma_{\mathrm{m}}k_{\mathrm{B}%
}T_{\mathrm{eff}}/4m\omega_{\mathrm{m}}^{2}$, $k_{\mathrm{B}}$ is the
Boltzmann's constant and $T_{\mathrm{eff}}$ is the effective noise temperature.

In the absence of optomechanical coupling $\Omega_{\mathrm{eff}}$ and
$\Gamma_{\mathrm{eff}}$ respectively represent the angular resonance frequency
detuning (with respect to $\omega_{\mathrm{m}}$) and the damping rate of the
isolated mechanical resonator, i.e. $\Omega_{\mathrm{eff}}=\omega_{0}=0$ and
$\Gamma_{\mathrm{eff}}=\gamma_{0}=\gamma_{\mathrm{m}}$. However, with a finite
optomechanical coupling both $\Omega_{\mathrm{eff}}$ and $\Gamma
_{\mathrm{eff}}$ are modified when the cavity is externally driven. Consider
the case where a monochromatic tone having normalized amplitude $a_{\mathrm{p}%
}$ and angular frequency $\omega_{\mathrm{p}}$ is injected into the feedline
in order to drive the microwave cavity. For this case the effective linear
angular frequency detuning $\omega_{0}$ and the effective linear damping rate
$\gamma_{0}$ become $\omega_{0}=2G^{2}E_{\mathrm{c}}\omega_{\mathrm{m}}%
^{-1}\operatorname{Im}\Xi_{1}$ and $\gamma_{0}=\gamma_{\mathrm{m}}%
+2G^{2}E_{\mathrm{c}}\omega_{\mathrm{m}}^{-1}\operatorname{Re}\Xi_{1}\left(
d,g\right)  $, respectively, where $\Xi_{l}\left(  d,g\right)  =\left[
-i\left(  d+l\right)  +g\right]  ^{-1}+\left[  -i\left(  d-l\right)
-g\right]  ^{-1}$, $g=\gamma_{\mathrm{c}}/\omega_{\mathrm{m}}$ is the
normalized cavity damping rate, and $E_{_{\mathrm{c}}}$ is the average number
of photons in the cavity in steady state, which is related to $a_{\mathrm{p}}$
by $E_{\mathrm{c}}=\left\vert a_{\mathrm{p}}\right\vert ^{2}\omega
_{\mathrm{m}}^{-2}\left(  d^{2}+g^{2}\right)  ^{-1}$ when optomechanical
coupling is disregarded \cite{Armour_440,Buks_454}. The optomechanical
coupling constant $G$ represents the shift in the effective cavity angular
resonance frequency induced by displacing the mechanical resonator by its zero
point amplitude.%

\begin{figure}
[ptb]
\begin{center}
\includegraphics[
height=2.5114in,
width=3.2396in
]%
{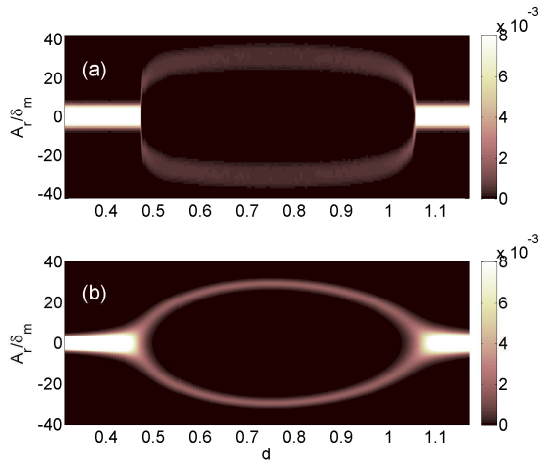}%
\caption{(color online) PSD in steady state as a function of the normalized
detuning $d$. The measured PSD seen in panel (a) is extracted using the
technique of state tomography. The calculated PSD in panel (b) is obtained
from the steady state solution of Eq. (\ref{Fokker-Planck}). The following
device parameters have been employed in the calculation:
$G=0.013\operatorname{Hz}$ (corresponding to frequency shift per displacement
of $55\operatorname{MHz}\operatorname{\mu m}^{-1}$) and $\gamma_{2}%
\delta_{\mathrm{m}}^{2}\gamma_{\mathrm{m}}^{-1}=2.0\times10^{-4}$. The rate of
nonlinear damping $\gamma_{2}$ is taken to be independent on the cavity
driving parameters since the optomechanical contribution to $\gamma_{2}$ is
found to be negligible small.}%
\label{Fig Steady State PSD}%
\end{center}
\end{figure}

In the current study we focus on the region close to the threshold of
self-excited oscillation, for which the effect of resonance frequency detuning
$\Omega_{\mathrm{eff}}$ can be disregarded. For such a case the Langevin
equation (\ref{A dot}) for the complex amplitude $A$ can be expressed in a
two-dimensional vector form as $\mathbf{\dot{A}}+\nabla\mathcal{H}%
=\mathbf{\xi}$ where $\mathbf{A}=\left(  A_{x},A_{y}\right)  $, $A_{x}$ and
$A_{y}$ are the real and imaginary parts of $A$, respectively, the noise term
is $\mathbf{\xi}=\left(  \xi_{x},\xi_{y}\right)  $, and the scalar function
$\mathcal{H}$ is given by $\mathcal{H}=\left(  \gamma_{0}/2\right)  A_{r}%
^{2}+\left(  \gamma_{2}/4\right)  A_{r}^{4}$ \cite{Hempstead_350,Baskin_563}.
The corresponding Fokker-Planck equation for the PSD $\mathcal{P}$ can be
written as \cite{Risken_Fokker-Planck}%
\begin{equation}
\frac{\partial\mathcal{P}}{\partial t}-\nabla\cdot\left(  \mathcal{P}%
\nabla\mathcal{H}\right)  -\Theta\nabla\cdot\left(  \nabla\mathcal{P}\right)
=0\;. \label{Fokker-Planck}%
\end{equation}

Consider the case where $\gamma_{2}>0$, for which a supercritical Hopf
bifurcation occurs when the linear damping coefficient $\gamma_{0}$ vanishes.
Above threshold, i.e. when $\gamma_{0}$ becomes negative the amplitude $A_{r}$
of self-excited oscillations is given by $A_{r0}=\sqrt{-\gamma_{0}/\gamma_{2}%
}$. The Fokker-Planck equation (\ref{Fokker-Planck}) can be used to evaluate
the normalized PSD function in steady state $W\left(  A\right)  $, which is
found to be given by $W=Z^{-1}e^{-\mathcal{H}/\Theta}$, where $Z$ is a
normalization constant (partition function)
\cite{Hempstead_350,Risken_Fokker-Planck,Yuvaraj_210403}. Note that $W$ is
independent on the angle $A_{\theta}$ of the complex variable $A$.

Panel (a) of Fig. \ref{Fig Steady State PSD} exhibits the measured PSD in
steady state as a function of the normalized detuning $d$ with a fixed pump
power of $P_{\mathrm{p}}=0.63%
\operatorname{\mu W}%
$. The PSD is extracted from the measured displacement of the mechanical
resonator, i.e. from the off-reflected signal of the optical cavity, using the
technique of state tomography \cite{Vogel_2847,Yuvaraj_210403}. Both the
measured PSD [panel (a)] and the calculated one $W=Z^{-1}e^{-\mathcal{H}%
/\Theta}$ [panel (b)] are plotted vs. the normalized radial amplitude
$A_{r}/\delta_{\mathrm{m}}$, where $\delta_{\mathrm{m}}=\sqrt{2\Theta
/\gamma_{\mathrm{m}}}$. The device parameters that have been employed for
generating the theoretical plot are listed in the figure caption. As $d$ is
increased from its initial value in the blue-detuning regime, the mechanical
element loses its stability due to the interaction with the microwave cavity
and undergoes a supercritical Hopf bifurcation at $d=0.474$. When $d$ is
further increased, the energy stored in the cavity is diminished and the
optomechanical coupling becomes smaller, until eventually the system regains
its stability at $d=1.06$. No hysteresis is found at this power level when $d$
is swept upwards and downwards. Note that the measured PSD [see panel (a)]
exhibits larger width around the average value [in comparison to theory, see
panel (b)] due to added measurement noise of displacement detection.%

\begin{figure}
[ptb]
\begin{center}
\includegraphics[
height=3.5717in,
width=3.2396in
]%
{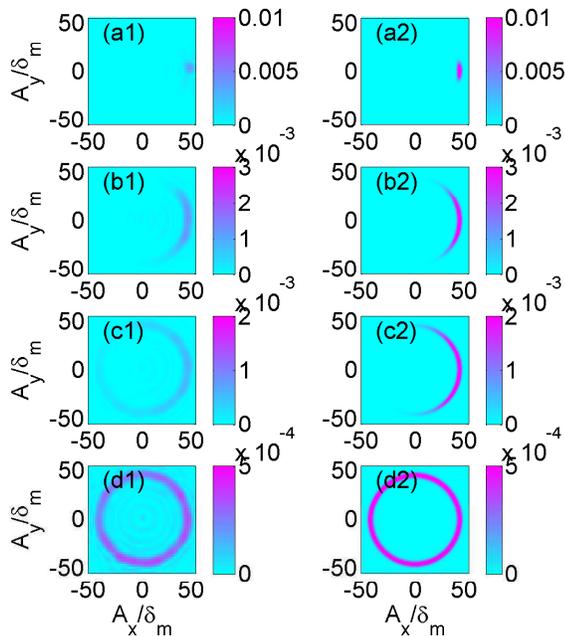}%
\caption{(color online) Measured (left) and calculated (right) PSD in steady
state self-excited oscillation. The normalized dwell time $\gamma_{\mathrm{m}%
}t_{\mathrm{d}}$ is $0.12$, $1.2$, $7.5$ and $12$ for the panels labeled by
the letters a, b, c, and d, respectively.}%
\label{Fig FP Ring}%
\end{center}
\end{figure}

The phase of self-excited oscillation in steady state randomly drifts in time
due to the effect of external noise. In addition, noise gives rise to
amplitude fluctuations around the average value $A_{r0}$. To experimentally
study these effects self-excited oscillation is driven by injecting a
monochromatic pump tone having normalized detuning $d=0.7$ and power
$P_{\mathrm{p}}=0.77%
\operatorname{\mu W}%
$ into the feedline. The off-reflected signal from the optical cavity is
recorded in two time windows separated by a dwell time $t_{\mathrm{d}}$. While
the data taken in the first time window is used to determined the initial
amplitude and phase of self-excited oscillation, the data taken in the second
one is used to extract PSD using tomography. The results are seen in Fig.
\ref{Fig FP Ring} for 4 different values of the dwell time $t_{\mathrm{d}}$.
While the left panels show the measured PSDs, the panels on the right exhibit
the calculated PSDs obtained by numerically integrating the Fokker-Planck
equation (\ref{Fokker-Planck}). For the longest dwell time the phase of
self-excited oscillation becomes nearly random, and consequently the PSD
becomes nearly independent on the angle $A_{\theta}$.

While cavity excitation in the measurements of PSD in steady state is
performed using a single microwave synthesizer, the time-resolved experiment,
in which the transient from cooling to heating is recorded, is performed using
two synthesizers. The first one, having negative normalized detuning
$d_{1}=-0.475$, serves during the cooling stage, whereas the second one,
having the opposite normalized detuning $d_{2}=-d_{1}$, is employed for
driving self-excited oscillation. Both synthesizers are connected to an RF
switch through two $-26$ dB directional couplers. The attenuated signal from
the couplers is connected to the input ports of a microwave switch, while the
through signal from the two synthesizers is mixed using an RF mixer. The IF
port of the mixer is used for triggering a pulse generator, which, in turn,
triggers the RF switch. Both the time delay between the two triggers and the
power levels of both synthesizers are carefully tuned in order to achieve a
smooth transition from cooling to heating. This is done by monitoring the
off-reflected signal shortly after switching and by minimizing signal ringing.
A smooth transition is achieved when both amplitude and phase of cavity fixed
point before switching coincide with the amplitude and phase corresponding to
the cavity fixed point after switching (which is unstable). Note that the
relative phase is a periodic function of the time delay with a period given by
the inverse frequency difference between the two synthesizers $\pi/d_{2}%
\omega_{\mathrm{m}}=1%
\operatorname{\mu s}%
$. The power level of both synthesizers is set close to $1.0%
\operatorname{\mu W}%
$ (with fine adjustments to eliminate ringing).

\begin{figure}
[ptb]
\begin{center}
\includegraphics[
height=2.5114in,
width=3.2396in
]%
{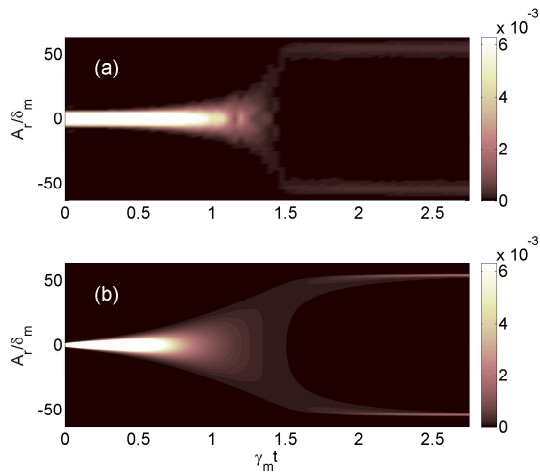}%
\caption{(color online) Time resolved PSD as a function of the normalized
elapsing time $\gamma_{\mathrm{m}}t$ since the switching from cooling to
heating. Panel (a) exhibits the measured PSD whereas the calculated PSD, which
is obtained by numerically integrating the Fokker-Planck equation
(\ref{Fokker-Planck}), is seen in panel (b).}%
\label{Fig TR PSD}%
\end{center}
\end{figure}

Fig. \ref{Fig TR PSD} shows the PSD as a function of the normalized delay time
$\gamma_{\mathrm{m}}t$, where $t$ is the elapsed time since the switching from
cooling to heating. For each measurement, the cavity is first excited with
normalized detuning $d_{1}$ for $5%
\operatorname{s}%
$, a time duration which is sufficiently long to reach steady state in the
cooling stage. The off reflected signal from the optical cavity is employed
for extracting the PSD using tomography. Panel (a) shows the measured PSD
whereas panel (b) shows the calculated PSD, which is obtained by numerically
integrating the Fokker-Planck equation (\ref{Fokker-Planck}). Note that the
PSD is expected to be independent on the angle $A_{\theta}$ of the complex
variable $A$ provided that the transition from cooling to heating is smooth.%

Analytical approximation to the time evolution of $\mathcal{P}$ that is
generated by the Fokker-Planck equation (\ref{Fokker-Planck}) can be obtained
for the case of short delay times $t>0$, for which $A_{r}$ is sufficiently
small to allow disregarding the effect of nonlinear damping. For this case,
$\mathcal{P}$ is found to be a Gaussian distribution given by $\mathcal{P}%
=\pi^{-1}\delta_{\mathrm{H}}^{-2}\exp\left(  -A_{r}^{2}/\delta_{\mathrm{H}%
}^{2}\right)  $, with exponentially growing width $\delta_{\mathrm{H}}$ given
by%
\begin{equation}
\delta_{\mathrm{H}}=\sqrt{\frac{2\Theta}{\gamma_{\mathrm{ba}}+\gamma
_{\mathrm{m}}}\left(  1+\frac{2\gamma_{\mathrm{ba}}\left(  e^{2\left(
\gamma_{\mathrm{ba}}-\gamma_{\mathrm{m}}\right)  t}-1\right)  }{\gamma
_{\mathrm{ba}}-\gamma_{\mathrm{m}}}\right)  }\;,\label{delta_H}%
\end{equation}
where $\gamma_{\mathrm{ba}}=2G^{2}E_{\mathrm{c}}\omega_{\mathrm{m}}%
^{-1}\operatorname{Re}\Xi_{1}\left(  d_{1},g\right)  $ is the back-action
contribution to the mechanical damping rate.

The process of PSD expansion, which is triggered by switching cavity detuning
from cooling to heating, can be employed under appropriate conditions for the
generation of a quantum superposition state \cite{Loerch_11015,Armour_440}.
Increasing the value of $\gamma_{\mathrm{ba}}$ by increasing the power
injected into the microwave cavity allows both, enhancing cooling efficiency
before switching, and accelerating PSD expansion after switching [see Eq.
(\ref{delta_H})]. The latter is highly desirable since experimental
observation of a superposition state is possible only if the time needed to
generate the state is shorter than the corresponding decoherence time. On the
other hand, while nonlinearity in the response of the microwave cavity has
been disregarded in the data analysis presented above, such effects may play
an important role at higher power levels
\cite{Dahm_2002,Suchoi_174525,Yurke_5054,Boerkje_53603,Nation_104516,Suchoi_1405_3467}%
. Taking full advantage of nonlinearity for enhancing the efficiency of
cooling \cite{Nation_104516} and for the suppression of decoherence
\cite{Buks_454} may open the way for experimental exploration of
non-classicality at a macroscopic scale in such systems.

This work was supported by the Israel Science Foundation, the bi-national
science foundation, the Security Research Foundation in the Technion, the
Israel Ministry of Science, the Russell Berrie Nanotechnology Institute and MAFAT.

\newpage
\bibliographystyle{apsrev}
\bibliography{acompat,Eyal_Bib}

\end{document}